\documentclass[cits]{PoS-pdf}

\usepackage{subfig} 
\usepackage{url}
\usepackage{wrapfig}
\usepackage{eufrak} 
\usepackage{amsmath}   
\usepackage{eurosym}   

\usepackage{mathptmx}
\usepackage{graphicx}
\usepackage[normalem]{ulem} 
\usepackage{color}
\let\normalcolor\relax 
\definecolor{myred}{rgb}{0.6,0,0} 
\definecolor{myblue}{rgb}{0,0.2,0.4}
\definecolor{mygreen}{rgb}{0,0.9,0.1}
\definecolor{Orange}{rgb}{1.,0.65,0.}
\usepackage{color}
\let\normalcolor\relax 
\definecolor{myred}{rgb}{1.0,0,0} 
\definecolor{mygreen}{rgb}{0,0.9,0.1} 
\definecolor{myblue}{rgb}{0,0.2,0.4}
\definecolor{mygray}{rgb}{.8,.8,.8}
\definecolor{darkorange}{rgb}{1, 0.549, 0}
\definecolor{purple}{rgb}{0.6,0.4,0.6}
\definecolor{mymagenta}{rgb}{0.6,0.4,0.6}
 \definecolor{LightCyan} {rgb}{0.88,1.,1.}
 \definecolor{Orange} {rgb}{1.,0.65,0.}
 \definecolor{PaleGreen} {rgb}{0.6,0.98,0.6}
 \definecolor{Pink} {rgb}{1.,0.75,0.8}
\definecolor{Red}{rgb}{1,0,0}
   \definecolor{Blue}{rgb}{0,0,1}
   \definecolor{Yellow}{rgb}{1,1,0}
   \definecolor{Orange}{rgb}{1,0.4,0}
   \definecolor{Pink}{rgb}{1,0,1}
   \definecolor{Purple}{rgb}{0.5,0,0.5}
   \definecolor{Teal}{rgb}{0,0.5,0.5}
   \definecolor{Navy}{rgb}{0,0,0.5}
   \definecolor{Aqua}{rgb}{0,1,1}
   \definecolor{Lime}{rgb}{0,1,0}
   \definecolor{Green}{rgb}{0,0.5,0}
   \definecolor{Olive}{rgb}{0.5,0.5,0}
   \definecolor{Maroon}{rgb}{0.5,0,0}
   \definecolor{Brown}{rgb}{0.6,0.4,0.2}
   \definecolor{Black}{gray}{0}
   \definecolor{Gray}{gray}{0.5}
   \definecolor{Silver}{gray}{0.75}
   \definecolor{White}{gray}{1}

\definecolor{darkblue}{rgb}{0,0,.5}

\usepackage{rotating}

\numberwithin{equation}{section}
\numberwithin{figure}{section}
\numberwithin{table}{section}

\newcommand{\be}{\begin{equation}}
\newcommand{\ee}{\end{equation}}
\newcommand{\bea}{\begin{eqnarray}}
\newcommand{\eea}{\end{eqnarray}}

\newcommand{\nl}{\nonumber \\}

\def \Yps {{\rm{Y}}}

\def \ol {\texttt{OLEC }{}}

\newcommand{\nn}{\nonumber}

\title{%
\small BI--TP 2012/46
\\
LPN 12--124
\\ 
\vspace*{1cm}
\LARGE
Efficient contraction of 1-loop N-point tensor integrals 
}

\ShortTitle{Contraction of N-point tensor integrals}

\author{\speaker{Jochem Fleischer}\\
        Fakult\"at f\"ur Physik, Universit\"at Bielefeld, Universit\"atsstr. 25,  33615
        Bielefeld, Germany
        \\
        E-mail: \email{Fleischer@physik.uni-bielefeld.de}
}

\author{ Janusz Gluza\\
Institute of Physics, University of Silesia, Uniwersytecka 4, PL-40007 Katowice, Poland
\\E-mail:
\email{janusz.gluza@us.edu.pl}}

\author{ Marek Gluza\\
AGH -- University of Science and Technology, Faculty of Physics and Applied Computer Science, al. A.
Mickiewicza, 30-059 Krak\'ow, Poland
}

\author{Tord Riemann\\
        Deutsches Elektronen-Synchrotron DESY, Platanenallee 6, 15738 Zeuthen, Germany
        \\
        E-mail: \email{Tord.Riemann@desy.de}
}

\author{Radomir Sevillano\\
Institute of Physics, University of Silesia, Uniwersytecka 4, PL-40007 Katowice, Poland
\\ E-mail: \email{radomir.sevillano-borkowski@us.edu.pl}       }

\abstract{%
A new approach for the reduction of tensor integrals is described.
The standard decomposition \`{a} la Davydychev is applied. 
Integrals with higher indices are then expressed in
terms of scalar higher-dimensional integrals with generic indices.
The approach allows to perform contractions with external momenta in a particularly efficient manner. 
This is due to the possibility to perform analytically the resulting sums over the indices of
 products of signed minors and scalar products of chords.
Advantages of this approach for the calculation of cross sections are described. 
We are preparing the numerical package
 \ol  (\texttt{O}ne \texttt{L}oop \texttt{E}xternal \texttt{C}ontractions) with interfaces in Mathematica for
algebraic and in C++ for numerical calculations. 
First numerical results are discussed. }

\FullConference{Loops and Legs in Quantum Field Theory - 11th DESY Workshop on Elementary Particle Physics\\
                 April 15-20, 2012\\
                 Wernigerode, Germany}

\begin{document}

\section{\label{Intro} Introduction}
Starting from the one-loop level, calculations of the squared amplitudes in Quantum Field Theory
bring us to consider $n$-point tensor integrals of rank $R$,   $(n,R)$-integrals
\begin{equation}
\label{definition}
 I_n^{\mu_1\cdots\mu_R} =  ~~\int \frac{d^d k}{i\pi^{d/2}}~~\frac{\prod_{r=1}^{R}
k^{\mu_r}}{\prod_{j=1}^{n}c_j^{\nu_j}}, 
\end{equation}
\\
where 
$d =  4-2 \epsilon$ and denominators $c_j$ have   {indices} $\nu_j$ and    {chords} $q_j$
\begin{equation}\label{propagators}
c_j = (k-q_j)^2-m_j^2 +i\varepsilon .
\end{equation}
There are many strategies how to solve such tensor integrals, see e.g.  
{\cite{Bern:2008ef,Binoth:2010ra,AlcarazMaestre:2012vp-2,Blumlein:2012lole}} { and references
therein.}.  

Our procedure and notations are described in { \cite{Fleischer:2010sq,Fleischer:2011nt} and in
references quoted therein}.
There are two main steps which are applied to (\ref{definition}).
  First,   higher dimensional scalar integrals are used. This is Davydychev's
decomposition {\cite{Davydychev:1991va}}, e.g. for a $R=2$ tensor with a general number of
external legs $n$
\begin{eqnarray}
\label{tensor2}
 I_{n}^{\mu\, \nu}& =& \int ^{d} k^{\mu} \, k^{\nu} \, \prod_{r=1}^{n} \, {c_r^{-1}} 
  =  \sum_{i,j=1}^{n} \, q_i^{\mu}\, q_j^{\nu} \, \nu_{ij} \,  \, I_{n,ij}^{[d+]^2} -\frac{1}{2}
   \,  g^{\mu \nu} \, I_{n}^{[d+]} .
\end{eqnarray}
We use the notation
\bea
[d+]^l=d+2l.
\eea
Second, recursive relations as derived in { \cite{Tarasov:1996br,Fleischer:1999hq} } bring the
expression into a simpler form, e.g. explicitly
for pentagons
\bea
\label{A522}
\nu_{ij} I_{5,ij}^{[d+]^2}
&=&
-\frac{{0\choose j}_5}{\left(  \right)_5} I_{5,i}
^{[d+]} +
 \sum_{s=1,s \ne i}^{5} \frac{{s\choose j}_5}{\left(  \right)_5}  I_{4,i}^{[d+],s} +
   \frac{{i\choose j}_5}{\left(  \right)_5} I_{5}^{[d+]}.
\eea
As may be seen, there is a decrease in both dimension and indices.
Finally, recursively, we go down to basic scalar integrals in higher dimension - but at  
the price of appearance of 
inverse Gram determinants $\left( \right)_5$ in (\ref{A522}). It is known  
that inverse Gram determinants can cause trouble in realistic physical applications, because there are
kinematical regions where they can be small or even vanish.

Here, the main aspect of the new approach comes into action: 
After insertion of (\ref{A522}) into (\ref{tensor2}), the chords $q_i,q_j$ are contracted with external
momenta (expressed as simple differences of chords $q_a, q_b$). 
One observes the appearance of ``auxiliary vectors''
\begin{eqnarray}
\label{av}
 Q_s^{\nu} &=& \sum_{j=1}^{5}  q_j^{\nu} \frac{{s\choose i}_5}
  {{\left(  \right)}_5}
,~~~ s=0, \ldots, 5. 
\end{eqnarray}
Contractions of the vectors with a chord $q_a$ immediately eliminate
the unwanted inverse of $\left( \right)_5$.
At first glance this is really surprising and we want to stress this property in particular.
If we would arrange cancellations of inverse Gram determinants $\left( \right)_5$ 
in (\ref{tensor2}) in a standard way, much more effort is needed 
\cite{Fleischer:2007ff,Diakonidis:2008ij,Fleischer:2010sq}.

To be more precise, let us give a manifest example, taken from subsection \ref{V1}:
\bea
(q_a \cdot Q_s) 
=
\sum_{j=1}^{n-1} (q_a  \cdot q_j) \frac{{s\choose j}_n}
{{\left( \right)}_n}
~~~=~~~\frac{1}{2}
\left({\delta}_{as}-{\delta}_{ns}\right),~~~ s=1, \ldots, 5..
\label{basic}
\eea
 The $Q$ vectors, when contracted with a chord in (\ref{basic}), give just Kronecker $\delta$-functions. The
inverse
Gram determinant  $\left( \right)_5$, present in intermediate steps, disappears automatically.  
 
Let us account now the main advantages of the new approach where chords are contracted with $Q$ vectors
(signed minors):
\begin{enumerate}
\item 
{\underline{Cancellation of inverse Gram determinants $\left( \right)_5$.}} 
\\
Explained as above
- occasionally also cancellation by subtraction.
\item 
{\underline{Elimination of the metric tensor $g_{\mu \nu}$.}}
\\
The integral (\ref{tensor2}) in its original form includes the $g_{\mu \nu}$ metric tensor. 
\\
The identity (valid in $d=4$ dimensions)
\bea
\frac{1}{2} g^{\mu\, \nu}=\sum_{i,j=1}^{5}  \frac{{i\choose j}_5}{\left(  \right)_5}\, q_i^{\mu}\, q_j^{\nu}
\label{gmvnv}
\eea
  introduces the inverse of $\left( \right)_5$ again. So, as argued before, it should cancel out. In
fact, 
e.g. for pentagons $(n=5)$,
\bea
\label{tensortwo}
  I_{5}^{\mu\, \nu}=
   \sum_{i,j=1}^{4} \, q_i^{\mu}\, q_j^{\nu} \, {\nu}_{ij} \,  \, I_{5,ij}^{[d+]
^2} -\frac{1}{2}
   \, g^{\mu \nu}  \, I_{5}^{[d+]} , 
\eea
and the last term in (\ref{A522}) cancels against the $g^{\mu \nu}$ term in (\ref{tensortwo}).
After the $g^{\mu \nu}$ cancellation  and the contractions with chords, the remaining inverse of 
$\left(\right)_5$, which was  produced in intermediate steps, disappears as well.
\item {\underline{Presence of Kronecker$\delta$-functions and simple kinematical factors $Y$.}} 
\\
This is probably the main feature of the approach. Due to relations like (\ref{basic}) and
 \bea
\label{Z1}
(q_a \cdot Q_0) &\equiv&\sum_{j=1}^{n-1} (q_a  \cdot q_j) \frac{{0\choose j}_n}
{{\left(\right)}_n}=-\frac{1}{2}\left(
 Y_{an}-Y_{nn} \right),
\eea
where $Y_{ij}=-(q_i-q_j)^2+m_i^2+m_j^2$ are simple kinematical functions, the expressions for 
contracted tensor integrals presented in the next section  take a simple form. Take  as an example the 
(\ref{reduc1}) from the next section. 
For given indices $a,b$ and $t$, most of the terms vanish.
\end{enumerate}
 The features are independent of multiplicity (i.e. the number of external legs) and rank of the
tensor integrals under consideration.

\section{\label{allow} The contractions
}
We consider now 5-point functions and eliminate the inverse 5-point Gram determinants $\left(\right)_5$.
Inverse 4-point Gram determinants ${s\choose s}_5$ will be treated separately.
The explicit formulae cover tensor integrals until rank $R=3$.
\subsection{The vector integral:~~ $I_5^{\mu}$}\label{V1}
The vector integral may be written in terms of the scalar 5-point function $E$ and 4-point scalar functions
$I_4^s$, where $s$ indicates the scratched line of the 5-point topology:
\bea
I_{5}^{\mu}=E Q_0^{\mu}-\sum_{s=1}^{5} I_4^s Q_s^{\mu},~~~~~~~~~~~~~~
E=\frac{1}{{0\choose 0}_5}\sum_{s=1}^{5} {s\choose 0}_5 I_4^s.
\eea
 For $q_n=0,~~ a=1, \dots , n-1,~~s=1, \dots n$, the multiplication with chords introduces \eqref{basic} and
\eqref{Z1},
and the final expression  for a contraction of the vector integral $I_5^{\mu}$ with an arbitrary chord
$q_{a,\mu}$ becomes
\bea
(q_a \cdot I_5)
&=&
 E ~(q_a \cdot Q_0)-\sum_{s=1}^5 I_4^s ~(q_a \cdot Q_s)
\nl
&=&-\frac{1}{2} E~ (Y_{a5}-Y_{55})-\frac{1}{2}
\sum_{s=1}^5 
I_4^s ~\left({\delta}_{as}-\delta_{5s} \right). \label{eqCE1}
\eea
Here we have applied \eqref{basic} and \eqref{Z1} for $n=5$. In fact no inverse Gram determinants
have appeared.

\subsection{The tensor integrals of rank $R=2$: ~~ $I_5^{\mu \nu}$} \label{V2}
From equation (3.5) of \cite{Fleischer:2010sq} one gets 
\bea
\label{twoteetwo}
I_{5}^{\mu\, \nu}=I_{5}^{\mu} Q_0^{\nu}-\sum_{s=1}^{5} I_4^{\mu,s} Q_s^{\nu},
\eea
with
\bea
I_4^{\mu ,s}&=& - \sum_{i=1}^4 q_i^{\mu} I_{4,i}^{[d+],s} = 
Q_0^{s,\mu} I_4^{s}-\sum_{t=1}^5 Q_t^{s,\mu} I_3^{st}.
\eea
There are two kinds of contractions to be considered: with the metric tensor or with two chords.
In principle, one may also meet in chiral theories contractions of the form
\bea
q^{a \alpha} q^{b \beta} ~ \epsilon_{\alpha\beta\mu\nu}   ~ I_5^{\mu \nu} .
\label{epsi2}
\eea
Proper organization of the matrix element evaluation will exclude the appearance of such terms since
cross sections are either helicity and spin summed and thus pure scalar quantities or written in
terms of helicity states. 
In the latter case, the scalar factors with the loop functions are also free of $\gamma_5$, and the
$\epsilon$ tensor might influence the matrix element basis.
Nevertheless, we add one rank $R=2$ example in order to demonstrate that this kind of sums may also be
treated. 

\subsubsection{Contractions with chords: $q_{a \mu} q_{b \nu}I_5^{\mu \nu}$}\label{V2a}
By contraction, \eqref{twoteetwo} rewrites into
\bea
\label{contr2}
q_{a \mu} q_{b \nu} I_5^{\mu \nu} =(q_a \cdot I_5) (q_b \cdot Q_0) 
 -\sum_{s=1}^{5} \left(q_a \cdot I_4^s \right)
(q_b \cdot Q_s).
\eea
Here no inverse $\left( \right)_5$ occurs anymore. Inserting the sums $\Sigma_a^{2,s}$ and 
$\Sigma_a^{1,st}$ from \cite{Fleischer:2011nt},  $\left(q_a \cdot I_4^s \right)$ can be written as
\bea
(q_a \cdot I_4^s )=\frac{1}{{s\choose s}_5} \left[
\Sigma_a^{2,s} I_4^s - \Sigma_a^{1,st} I_3^{st} \right] = -\frac{1}{2}  \left\{(Y_{a5}-Y_{55}) I_4^s+
\sum_{t=1}^5 ({\delta}_{at}- {\delta}_{5t}) I_3^{st} + ({\delta}_{as}- {\delta}_{5s}) R^s \right\},
\label{DSig}
\eea
where we introduced the abbreviation 
\bea
R^s \equiv
\frac{1}{{s\choose s}_5} \left[{s\choose 0}_5 I_4^s - \sum_{t=1}^5 {s\choose t}_5 I_3^{st} \right]=
\frac{1}{{0s\choose 0s}_5} \left[{s\choose 0}_5 I_4^{[d+],s} - \sum_{t=1}^5 {0s\choose 0t}_5 I_3^{st} \right].
\label{Rs}
\eea
The $2^{nd}$ representation of $R^s$  in \eqref{Rs} can be used if the four-point Gram determinant ${s\choose
s}_5$ is small or even vanishes.
\subsubsection{Contractions with the metric tensor: $g_{\mu \nu} I_5^{\mu \nu}$}\label{V2b}
The second scalar which can be 
constructed from the tensor of rank 2 is $g_{\mu \nu} I_5^{\mu \nu}$. Due to
\eqref{twoteetwo} we have to evaluate the following scalar products:
 \bea
(Q_0 \cdot Q_0)&=&\frac{1}{2}\left[\frac{{0\choose 0}_5}{\left(  \right)_5}+Y_{55} \right], \nn \\
(Q_0 \cdot Q_s)&=&\frac{1}{2}\left[\frac{{s\choose 0}_5}{\left(  \right)_5}-{\delta}_{s5} \right], \nn \\
(Q_0^s \cdot Q_s)&=&-\frac{1}{2}{\delta}_{s5}, \nn \\
(Q_t^s \cdot Q_s)&=&~~~0.
\label{scalpr}
\eea
In this case the terms with $\frac{1}{\left(  \right)_5}$ cancel and, not surprisingly, the result finally is
\bea
\label{Nosurprise}
g_{\mu \nu} I_5^{\mu \nu}=\frac{Y_{55}}{2} E + I_4^5.
\eea
\subsubsection{Contractions with the antisymmetric tensor: 
$q^{a \mu} q^{b \nu} \epsilon_{\mu\nu\alpha\beta} I_5^{\alpha \beta}$
}
A reduction of pseudoscalar contractions leads to expressions of the type 
\bea
P[C] = q_a^{\rho} q_b^{\lambda}~ {\epsilon}_{\mu \nu \rho \lambda} ~\sum_{i,j=1}^4  q_i^{\mu} q_j^{\nu}
C_{ij} .
\eea
This may be evaluated as follows:
\bea
P[C]
&=&
q_a^{\rho} q_b^{\lambda} ~{\epsilon}_{\mu \nu \rho \lambda} ~g_{~~{\mu}'}^{\mu}~ g_{~~{\nu}'}^{\nu}
\sum_{i,j=1}^4  q_i^{{\mu}'} q_j^{{\nu}'} C_{ij} 
\nn \\
&=&
\frac{4}{{\left( \right)}_5^2} q_a^{\rho} q_b^{\lambda}{~ \epsilon }_{\mu \nu \rho \lambda} 
\sum_{k,l=1}^4  {k\choose l}_5
q_k^{\mu} ~q_{l,{\mu}'}  \sum_{m,n=1}^4  {m\choose n}_5
q_m^{\nu}~ q_{n,{\nu}'}  \sum_{i,j=1}^4  q_i^{{\mu}'}q_j^{{\nu}'} C_{ij} 
\nn \\
&=&
\frac{4}{{\left( \right)}_5^2}
\sum^4_{\substack{ k,m=1 \\ a+b+k+m=10}}
\left[ q_a^{\rho} q_b^{\lambda}
{~ \epsilon }_{\mu \nu \rho \lambda}q_k^{\mu} q_m^{\nu} \right]   \sum_{l,n=1}^4  {k\choose l}_5 {m\choose
n}_5   \sum_{i,j=1}^4 (q_l   \cdot q_i) (q_n \cdot q_j)C_{ij} 
\nn \\
&=&
\frac{4}{{\left( \right)}_5^2} \det \left[q_1,q_2,q_3,q_4 \right]
\sum_{\substack{ k,m=1 \\ a+b+k+m=10}}^4 S(a,b,k,m) 
\sum_{l,n=1}^4  {k\choose l}_5 {m\choose n}_5  {\Sigma}_{ln}[C],
\eea
where $S(a,b,k,m)$ gives the sign of permutations needed to place the indices in increasing order. Further we
introduced the abbreviation
\bea
{\Sigma}_{ln}[C]=\sum_{i,j=1}^4 (q_l \cdot q_i) (q_n \cdot q_j)C_{ij},
\eea
the calculation of which has to be performed for specific cases. 

Let us choose for demonstrational purposes $C_{ij}={is\choose js}_5$:
\bea\label{Sigmaln}
{\Sigma}_{ln}[C]
&=&
\frac{1}{2} (q_l \cdot q_n ) {s\choose s}_5 - \frac{1}{4} {\left( \right)}_5
({\delta}_{ls}-{\delta}_{5s})({\delta}_{ns}-{\delta}_{5s}) 
 \\
&=&
\begin{cases}
\frac{1}{2} (q_l \cdot q_n ) {s\choose s}_5 - \frac{1}{4} {\left( \right)}_5 {\delta}_{ls} {\delta}_{ns}
&{\rm s} \ne 5 ,
\nn \\
\frac{1}{2} (q_l \cdot q_n ) {s\choose s}_5 - \frac{1}{4} {\left( \right)}_5 
& {\rm s} = 5.
\end{cases}
\eea
Next we have to evaluate the sum $\sum_{l,n=1}^4  {k\choose l}_5 {m\choose n}_5 \cdot {\Sigma}_{ln}[C]$.
Taking the first term in \eqref{Sigmaln}, to begin with, we have
\bea
\label{eq217}
\frac{1}{2}\sum_{l,n=1}^4 (q_l \cdot q_n ) {k\choose l}_5 {m\choose n}_5
&=&
\frac{1}{2}\sum_{l=1}^4{k\choose
l}_5\sum_{n=1}^4
(q_l \cdot q_n ) {m\choose n}_5 
\nn \\
&=&
\frac{1}{4}\sum_{l=1}^4{k\choose l}_5 {\delta}_{ml} {\left( \right)}_5
\nn \\
&=&
\frac{\left( \right)_5}{4}{k\choose
m}_5
,
\eea
and for $s\ne 5$ we thus have
\bea
\sum_{l,n=1}^4  {k\choose l}_5 {m\choose n}_5 \cdot {\Sigma}_{ln}[C]
&=&
\frac{\left( \right)_5}{4}\left[{k\choose
m}_5{s\choose s}_5-
{k\choose s}_5{m\choose s}_5 \right]
\nn \\
&=&
\frac{\left( \right)_5^2}{4} {ks\choose ms}_5.
\label{Resne5}
\eea
Similarly we have for $s=5$
\bea
\sum_{l,n=1}^4  {k\choose l}_5 {m\choose n}_5 \cdot {\Sigma}_{ln}[C]=\frac{\left( \right)_5}{4}\left[{k\choose
m}_5{5\choose 5}_5-
\sum_{l,n=1}^4{k\choose l}_5{m\choose n}_5 \right].~~~~~~~~~~~~~~~
\eea
Due to
\bea
\sum_{l=1}^5{k\choose l}_5=0,~~~{\rm i.e. } \sum_{l=1}^4{k\choose l}_5=-{k\choose 5}_5,
\eea
we get for $s=5$ the same result as in \eqref{Resne5}. 

Thus the total result reads for the specific choice of $C_{ij})$:
\bea
 P[{is\choose js}_5] = \det \left[q_1,q_2,q_3,q_4 \right] \sum_{\substack{k,m=1 \\a+b+k+m=10}}^4 S(a,b,k,m)
{ks\choose ms}_5.
\label{Fin}
\eea
In the summation of \eqref{Fin} all indices $a,b,k,m$ must be different, i.e. the sum contains only $2$ terms,
with
$k$ and $m$ exchanged. Since these terms come with opposite sign and \eqref{Resne5} is symmetric in $k$ and
$m$,
the final result for the present example vanishes, i.e.
\bea
P[{is\choose js}_5] 
&=&
 q_a^{\rho} q_b^{\lambda} {~ \epsilon }_{\mu \nu \rho \lambda} \sum_{i,j=1}^4  q_i^{\mu}
q_j^{\nu} {is\choose js}_5
\nn \\
&=&0.
\eea

\subsection{The tensors of rank $R=3$: ~~$I_5^{\mu \nu \lambda}$}\label{V3}
From eq.(3.5) of \cite{Fleischer:2010sq} one gets
\bea\label{T53}
I_5^{\mu \nu \lambda}=I_5^{\mu\nu} \cdot Q_0^{\lambda} -\sum_{s=1}^{5} I_4^{\mu\nu, s} \cdot Q_s^{\lambda},
\eea
where according to (3.12) in \cite{Fleischer:2010sq} it is
\bea
I_4^{\mu \nu,s}&=&I_4^{\mu,s} Q_0^{s,\nu} 
-\sum_{t=1}^5 I_3^{\mu , st} Q_t^{s,\nu}  -\frac{{\left( \right)}_5}{{s\choose s}_5}Q_s^{\mu} Q_s^{\nu} I_4^{[d+],s} , \nn \\
I_3^{\mu ,st}&=& - \sum_{i=1}^4 q_i^{\mu} I_{3,i}^{[d+],st} = 
Q_0^{st,\mu} I_3^{st}-\sum_{u=1}^5 Q_u^{st,\mu} I_2^{stu}.
\label{I4munu}
\eea

\subsubsection{Contractions with chords: ~~~$ q_{a \mu} q_{b \nu} q_{c \lambda} I_5^{\mu \nu
\lambda}$}\label{V2a3}
The contraction of \eqref{T53} with $q_{c \lambda}$ is trivial due to \eqref{basic} and
\eqref{Z1}. 
Further, the term $ q_{a \mu} q_{b \nu}I_5^{\mu \nu}$ is known from subsection \ref{V2a}.
The only new object to be investigated is therefore 
\bea
q_{a \mu} q_{b \nu} I_4^{\mu \nu,s}&&=\left(q_a \cdot I_4^s \right)\left(q_b \cdot Q_0^s \right)-
\sum_{t=1}^5\left(q_a \cdot I_3^{st} \right)\left(q_b \cdot Q_t^s \right)-\frac{{\left( \right)}_5}{{s\choose s}_5}
(q_a \cdot Q_s)(q_b \cdot Q_s) I_4^{[d+],s}.
\label{PRD1}
\eea 
New sums may be expressed by those derived in \cite{Fleischer:2011nt}: 
\bea
\label{contrstu}
(q_b \cdot Q_0^{s}) = \frac{1}{{s\choose s}_5} \Sigma_b^{2,s}, 
 \\
(q_b \cdot Q_t^{s}) = \frac{1}{{s\choose s}_5} \Sigma_b^{1,st},
\eea
and
\bea
(q_a \cdot I_3^{st} )
=
&&\frac{1}{{st\choose st}_5} \left[\Sigma_a^{3,st} I_3^{st} -\sum_{u=1}^5 \Sigma_a^{1,stu} I_2^{stu}\right] 
\nn \\
=&&-\frac{1}{2} \left\{(Y_{a5}-Y_{55})I_3^{st}+
\sum_{u=1}^5 ({\delta}_{au}- {\delta}_{5u}) I_2^{stu} +({\delta}_{as}- {\delta}_{5s})R^{st}+({\delta}_{at}- {\delta}_{5t})R^{ts}
\right\} ,
\label{repeat}
\eea
where we introduced as a further abbreviation
\bea
R^{st} 
&\equiv&
\frac{1}{{st\choose st}_5} \left[{st\choose 0t}_5 I_3^{st}-\sum_{u=1}^5 {st\choose ut}_5 I_2^{stu} \right]
\nl
&=&
\frac{1}{{0st\choose 0st}_5}\left[{st\choose 0t}_5 (d-2) I_3^{[d+],st}-\sum_{u=1}^5 {0st\choose 0ut}_5 I_2^{stu} \right].
\label{Rst}
\eea  
From
\bea
\Sigma_b^{2,s}&&=-\frac{1}{2} \left\{{s\choose s}_5 (Y_{b5}-Y_{55})+{s\choose 0}_5 ({\delta}_{bs}-
{\delta}_{5s}) \right\} 
,
 \\
\Sigma_b^{1,st}&&=~~~\frac{1}{2} \left\{{s\choose s}_5 ~({\delta}_{bt}- {\delta}_{5t})-{s\choose t}_5 ({\delta}_{bs}- {\delta}_{5s}) \right\},
\label{Deltas}
\eea
multiplying \eqref{DSig} and \eqref{repeat}, respectively, we see that the $({\delta}_{bs}-
{\delta}_{5s})$-terms in 
\eqref{Deltas} combine with the first terms in \eqref{DSig} and \eqref{repeat} to yield the contribution
\bea
\frac{1}{4} (Y_{a5}-Y_{55})({\delta}_{bs}-{\delta}_{5s})  R^s.
\eea
Collecting all contributions, the following relation is useful in order to to eliminate some
$\frac{1}{{s\choose
s}_5}$-terms:
\bea
R^{st}+\frac{{s\choose t}_5}{{s\choose s}_5}R^{ts}=\frac{{s\choose 0}_5}{{s\choose s}_5} I_3^{st}-\sum_{u=1}^5 
\frac{{s\choose u}_5}{{s\choose s}_5} I_2^{stu} .
\eea
The result for \eqref{PRD1} is explicitly symmetric in the indices $a,b$:
\bea
q_{a \mu} q_{b \nu} I_4^{\mu \nu,s}=q_{a \mu} q_{b \nu} {\overline I}_4^{\mu \nu,s}-\frac{1}{4}
({\delta}_{as}-{\delta}_{5s})({\delta}_{bs}-{\delta}_{5s}) \frac{{\left( \right)}_5}{{s\choose s}_5}
I_4^{[d+],s},
\label{reduc0}
\eea
and the first term on the right hand side is
\bea
q_{a \mu} q_{b \nu} {\overline I}_4^{\mu \nu,s}
&=&\frac{1}{4} (Y_{a5}-Y_{55})(Y_{b5}-Y_{55}) \cdot I_4^s\nn \\
&&+~\frac{1}{4} ({\delta}_{as}-{\delta}_{5s})({\delta}_{bs}-{\delta}_{5s}) \frac{1}{{s\choose s}_5}
\left[{s\choose 0}_5 R^s- \sum_{t=1}^5 {s\choose t}_5 R^{st} \right]\nn \\
&&+~\frac{1}{4}
\left[({\delta}_{as}-{\delta}_{5s})(Y_{b5}-Y_{55})+(Y_{a5}-Y_{55})({\delta}_{bs}-{\delta}_{5s}) \right] R^s
\nn \\
&&+~\frac{1}{4} \sum_{t=1}^5 \left[({\delta}_{as}-{\delta}_{5s})({\delta}_{bt}-{\delta}_{5t})+
({\delta}_{bs}-{\delta}_{5s})({\delta}_{at}-{\delta}_{5t}) \right] R^{st} \nn \\
&&+~\frac{1}{4} \sum_{t=1}^5
\left[({\delta}_{at}-{\delta}_{5t})(Y_{b5}-Y_{55})+(Y_{a5}-Y_{55})({\delta}_{bt}-{\delta}_{5t}) 
\right]I_3^{st} \nn \\
&&+~\frac{1}{4} \sum_{t=1}^5({\delta}_{at}-{\delta}_{5t}) ({\delta}_{bt}-{\delta}_{5t}) R^{ts}
+\frac{1}{4}\sum_{t,u=1}^5({\delta}_{bt}-{\delta}_{5t})({\delta}_{au}-{\delta}_{5u})I_2^{stu}.
\label{reduc1}
\eea
It is worth mentioning that the second term in \eqref{reduc0} may be written as
\bea
{\left( \right)}_5 I_4^{[d+],s}={0\choose 0}_5 I_4^s-\sum_{t=1}^5 {t\choose 0}_5 I_3^{st}-{s\choose 0}_5 R^{s}.
\eea
The $R^{s}$, $R^{st}$, $\sum_{t=1}^5 {s\choose t}_5 R^{st}$ and the scalar integrals in \eqref{reduc1}
are independent of the indices $a,b,c$ and can be considered as buildung blocks. The 
summation over $s,t,u$ is simplified by the explicit appearance of the Kronecker $\delta$-functions, i.e. 
all contractions are just given in terms of different combinations of the buildung blocks.

There still occur contributions with factor $\frac{1}{{s\choose s}_5}$ in \eqref{reduc0} and  in
\eqref{reduc1}. At first we observe that this
contribution is indeed quite ``localized'': in general one will be able to avoid the small ${s\choose s}_5$
to appear on line $s=5$ by choosing the integration momenta properly. Let say ${s_0\choose s_0}_5$ is small. 
Then the critical contribution will only occur if $a=b=c=s_0$, i.e. only in a single contraction. 

Nevertheless, if one wants to avoid the $\frac{1}{{s\choose s}_5}$ completely, one can express
the corresponding
contribution 
for the sum of these terms in \eqref{reduc0} and \eqref{reduc1}
in terms of higher dimensional integrals, making again use of  \eqref{Rs}. 
Let us call the sum $J_4^s$:
\bea
\label{J4s}
J_4^s &\equiv& 
\frac{1}{{s\choose s}_5}
\left\{ -{\left( \right)}_5 I_4^{[d+],s}+{s\choose 0}_5 R^s- \sum_{t=1}^5 {s\choose t}_5 R^{st} \right\} 
\\
&=&\frac{-1}{{0s\choose 0s}_5} \left\{ \left( \right)_5 (d-2)(d-1) I_4^{[d+]^2,s}-{0\choose 0}_5 I_4^{[d+],s}
+\sum_{t=1}^5 {t\choose 0}_5 (d-2) I_3^{[d+],st} +\sum_{t=1}^5 {0s\choose 0t}_5 R^{st} \right\}. \nn
\eea
The higher dimensional integrals can then be evaluated as described in \cite{Fleischer:2010sq}.

Let us call the second term on the right hand side of \eqref{T53} $C_{5,abc}$.
It
becomes, when contracted with momenta
$q_a,q_b$ and
$q_c$: 
\bea
C_{5,abc}&=& - ~ q_{a \mu} q_{b \nu} q_{c \lambda}
\sum_{s=1}^{5} I_4^{\mu\nu, s} \cdot Q_s^{\lambda}
\nl 
&=&-~ \sum_{s=1}^5 q_{a \mu} q_{b \nu} I_4^{\mu \nu,s} \cdot \frac{1}{2}
\left({\delta}_{cs}-{\delta}_{5s} \right).
\eea
The last term again contains \eqref{reduc0}.
Introducing ${\Yps}_a =Y_{a5}-Y_{55}$ and making use of \eqref{J4s}, we may collect terms and get
\bea
C_{5,abc}=\frac{1}{8} &&\left\{J_4^5-{\delta}_{ab}{\delta}_{ac} J_4^a +
{\Yps}_a {\Yps}_b \left(I_4^5-I_4^c \right) 
~~+{\Yps}_a \left(I_3^{b5}+I_3^{c5}-I_3^{bc}\right)+{\Yps}_b\left( I_3^{a5}+I_3^{c5}-I_3^{ac}\right) \right. \nn \\
&&\left.~~ +I_2^{ab5}+I_2^{ac5}+I_2^{bc5}-I_2^{abc} 
~~ -{\Yps}_a \left( R^5+{\delta}_{bc} R^b\right)-
{\Yps}_b \left( R^5+{\delta}_{ac} R^a\right)\right. \nn \\
&&\left.~~ -R^{5a}-R^{5b}-R^{5c} 
~~ +{\delta}_{ab}\left(R^{a5}-R^{ac} \right)+{\delta}_{ac} \left(R^{a5}-R^{ab}\right) ~~ +{\delta}_{bc}\left(R^{b5}-R^{ba}\right)
\right\}.
\nl
\eea

\subsubsection{Contractions with the metric tensor:~~~$g_{\mu \nu} q_{a \lambda} I_5^{\mu \nu \lambda}$  }
In order to calculate $g_{\mu \nu} I_5^{\mu \nu \lambda}$, we need $g_{\mu \nu} I_4^{\mu \nu }$ and thus
the further scalar products, see \eqref{I4munu}:
\bea
(Q_0^s \cdot Q_0^s)&=&\frac{1}{2 {s\choose s}_5}\left[{0s\choose 0s}_5+2 {s\choose 0}_5 {\delta}_{s5}\right]+
                       \frac{1}{2} Y_{55} , \nn \\               
(Q_s \cdot Q_s)&=&\frac{1}{2}\frac{{s\choose s}_5}{\left(  \right)_5}, \nn \\
(Q_t^s \cdot Q_0^s)&=&\frac{1}{2 {s\choose s}_5}\left[{ts\choose 0s}_5-{s\choose s}_5 {\delta}_{t5}+{s\choose t}_5 {\delta}_{s5}\right],
\nn \\
(Q_t^s \cdot Q_0^{st})&=&\frac{1}{2 {s\choose s}_5}\left[~~~~~~~~~~~-{s\choose s}_5 {\delta}_{t5}+{s\choose t}_5 {\delta}_{s5}\right],
\nn \\
(Q_t^s \cdot Q_u^{st})&=&0, 
\eea
which yields
\bea
\label{nottriv}
g_{\mu \nu} I_4^{\mu \nu, s }=
\frac{ Y_{55}}{2} I_4^s + I_3^{s5} +{\delta}_{s5} R^s
\eea
and finally
\bea
\label{twocon}
g_{\mu \nu} q_{a \lambda} I_5^{\mu \nu \lambda}=
-\frac{Y_{55}}{4} \left[(Y_{a5}-Y_{55}) E +I_4^a-I_4^5 \right]-\frac{1}{2} \left[(Y_{a5}-Y_{55}) I_4^5 +
I_3^{a5}-R^5 \right] .
\eea
It is remarkable that \eqref{nottriv} is trivial again for $s \ne 5$. For $s=5$, however, the standard
cancellation
of propagators does not work and for this case \eqref{nottriv} is indeed a useful result.

\section{The OLEC package -- first numerical results}
The relations introduced in the  previous sections have been implemented in Mathematica. A corresponding
file OLECv0.9.m can be found at \cite{olec-project-09}. 
Certainly, computer algebra programs like Mathematica are not the optimal and efficient tools for
ultimate numerical applications, they are slow. 
However, there are some reasons to start with Mathematica.
First, standard tensor reductions for five and six point functions discussed in \cite{Diakonidis:2008ij} have
been also implemented in the Mathematica package hexagon.m \cite{hexagon-project-10,Gluza:2009mj}.
The new method of direct
contractions of tensors as introduced in \cite{Fleischer:2011nt,Fleischer:2010sq} and
presented here can be compared with hexagon.m.
Second, if basic scalar integrals are known analytically, then the complete reductions are automatically given
also in an analytic form. This approach could be used for instance to investigate infrared structure of
amplitudes \cite{Gluza:2007uw,Gluza:2008tk,Diakonidis:2008dt,Diakonidis:2008ij}.
Of course, for numerical applications one will prefer packages written in Fortran or C++. 
For efficient reductions of five-point tensor integrals, see V. Yundin's C++ package PJFry
\cite{Fleischer:2011zz,Fleischer:2012et,yundin-phd-2012--oai:export}).

For the \emph{contracted} 5-point functions with which we are dealing here, the following notation is used:
\verb+CEx(chords)+, where "C" stands for "contracted" and "E" stands traditionally for pentagons. The
argument "chords" represents a list of indices, and their length depends on the rank "x" of the considered
object.
In this way, $CE1(a)$ is equivalent to (\ref{eqCE1}), $CE2(a,b)$ is equivalent to (\ref{contr2}) and
$CE3(a,b,c)$ is equivalent to appropriate relations given in subsection \ref{V2a}. For the objects contracted
with the metric tensor $g_{\mu \nu}$ we have two objects up to rank 3: 
  $CE2g$ and $CE3g(a)$. They correspond to (\ref{Nosurprise}) and (\ref{twocon}), respectively.

Now, we can compare these objects with corresponding objects which come from standard reductions. 
Below we will use the notation 
of LoopTools/FF \cite{Hahn:1998yk,Hahn:2010zi} for coefficients of corresponding tensorial objects $E0i$:
\begin{eqnarray}\label{eqCE3} 
CE1(a) &\equiv& q^{\mu_1}_a   \left\{ 
\sum\limits_{i=1}^4 q_{\mu_1}^i E0i[ee{i},invs]  \right\} ,
\\
CE2(a,b)
&\equiv& q^{\mu_1}_a q^{\mu_2}_b  \left\{ 
\sum\limits_{\{i,j\}=1}^4 q_{\mu_1}^i  q_{\mu_2}^j E0i[ee{ij},invs]  
+   g_{\mu_1 \mu_2}  E0i[ee{00},invs] \right\} \label{eqCE2} ,
\\
CE3(a,b,c)&\equiv& q^{\mu_1}_a q^{\mu_2}_b q^{\mu_3}_c \left\{ 
\sum\limits_{\{i,j,k\}=1}^4 q_{\mu_1}^i q_{\mu_2}^j q_{\mu_3}^j E0i[ee{ijk},invs] \right. 
\nonumber \\
&&+~ \left. \sum\limits_{i=1}^4 \left( g_{\mu_1 \mu_2} q_{\mu_3}^i+g_{\mu_1 \mu_3} q_{\mu_2}^i +
g_{\mu_2 \mu_3} q_{\mu_1}^i \right) E0i[ee{00i},invs]
\right\} .
\end{eqnarray}
For contractions with the metric tensor:
\begin{eqnarray}
CE2g &\equiv&  E0i[ee001,invs]  , \\
\nonumber \\
CE3g(a) &\equiv&  g^{\mu_1 \mu_2} q^{\mu_3}_a \left\{ 
\sum\limits_{\{i,j,k\}=1}^4 q_{\mu_1}^i q_{\mu_2}^j q_{\mu_3}^j E0i[ee{ijk},invs] \right. \nonumber \\
&+& \left. \sum\limits_{i=1}^4 \left( g_{\mu_1 \mu_2} q_{\mu_3}^i+g_{\mu_1 \mu_3} q_{\mu_2}^i+
g_{\mu_2 \mu_3} q_{\mu_1}^i \right) E0i[ee{00i},invs]
\right\} .
\end{eqnarray}
 Numerical results are shown in the file \verb+Contracts_examples.nb+.
 They are compared for various chord indices using the Looptools/FF and OneLoop \cite{vanHameren:2010cp}
packages.

Below we tabularize an example of results extracted from \verb+Contracts_examples.nb+ 
for kinematical points defined by:
\\ 
$p_1^2 = p_2^2 = 0;\; p_3^2 = p_5^2 = 49/256;\; p_4^2 = 9/100;\; 
\\
s_{12} = 4;\; s_{23} = -1/5;\;
s_{34} = 1/5;\; s_{45} = 3/10;\;
s_{15} = -1/2;\;
\\
m_1^2 = m_2^2 = m_3^2 = 49/256;\;
m_4^2 = m_5^2 = 81/1600.$ 
\begin{verbatim}
In: 
Do[Print[i, ") LoopTools: ", Rank1LoopTools[i], ", Hexagon: ", 
  Rank1hexagon[i], ", Contracts: ", CE1[i]], {i, 1, 5}]
Out:  
1) LoopTools: -19.1597+12.6772 I, Hexagon: -19.1597+12.6772 I, 
								Contracts: -19.1597+12.6772 I
2) LoopTools: -80.4038+91.9442 I, Hexagon: -80.4038+91.9442 I, 
								Contracts: -80.4038+91.9442 I  
3) LoopTools: -19.1525+12.4364 I, Hexagon: -19.1525+12.4364 I, 
								Contracts: -19.1525+12.4364 I
4) LoopTools: -17.496+10.67060 I, Hexagon: -17.496+10.67060 I, 
								Contracts: -17.496+10.67060 I
5) LoopTools: 0, Hexagon: 0. +0. I, Contracts: 0
  \end{verbatim}
      In addition, comparisons concerning the speed of calculation have been made. As expected, the 
Mathematica package OLEC.m v.0.9  is at least one order of magnitude faster than hexagon.m but still slower
than the Fortran package LoopTools/FF.
   
Thus, a second package OLEC v.0.9 written in C++ has been prepared for numerical studies. 
It should be assumed as the skeleton for the first version of the ultimate package. It has to be further
optimized and   additional refinements for specific kinematical points are needed, similar to those
which were implemented for the standard reductions within the PJFRY package. 
   
The OLEC package can be downloaded from \cite{olec-project-09}. It is linked with both the LoopTools/FF
and the OneLoop libraries.
 Our first package, where the contractions replace the explicit tensor reductions  
is at the moment comparable in speed to  numerical results of 5-point tensor reductions implemented in
LoopTools/FF.
This has been tested for specific rank 2 tensors in (\ref{eqCE2}) and  tensors of rank 3 in (\ref{eqCE3}). 
Details can be found in the module \verb+example_2+ of the Makefile in \cite{olec-project-09} where $CE2(1,2)$
and $CE3(1,2,3)$ can be calculated and compared with corresponding LoopTools objects using different
flags, for users
convenience. 
The result is not too surprizing.
Evidently, for higher tensor ranks the advantage of contractions compared to rank reductions, introducing
multiple sums,  should become more pronounced.

\section{Conclusions and outlook}

A new method of evaluation of contracted tensor integrals has been presented for five-point one loop
integrals.
So far individual objects like $CE1$, $CE2$ and $CE3$ have been resolved and implemented into first codes.
What remains is to apply it to the calculation of whole amplitudes for some physical processes. 

In the near future the first version 1.0 of the complete code in C++ and Fortran~90 is planned to be released.
Here we have shown the backbone of the code with first numerical estimations. In addition, higher than rank 3
tensor structures for 5-point functions and tensorial integrals  with more  than five external legs are in
preparation.  

\section*{Acknowledgments}
We would like to thank Andreas van Hameren, Thomas Hahn and Krzysztof Kajda for useful discussions and help.
Work supported by European Initial Training Network LHCPHENOnet
PITN-GA-2010-264564 and by Polish Ministry of Science and Higher Education from budget for
science for 2010-2013 under grant number N N202 102638.
  
%

\providecommand{\href}[2]{#2}\begingroup\raggedright\endgroup

\end{document}